\documentclass[11pt,twoside]{article}
\usepackage{asp2010}
\usepackage{amssymb}
\usepackage{graphicx,mathptm}
\newcommand{\be}{\begin{equation}}
\newcommand{\ee}{\end{equation}}
\newcommand{\bea}{\begin{eqnarray}}
\newcommand{\eea}{\end{eqnarray}}

\resetcounters

\begin{document}

\title{Reconnection Diffusion, Star Formation and Numerical Simulations}
\author{ A. Lazarian
\affil{Astronomy Department, University of Wisconsin, Madison, WI 53706, US}}

\begin{abstract}
We consider fast magnetic reconnection that takes place within turbulent magnetic flux
and show that the process results in diffusion of magnetic fields and matter, which we term
reconnection diffusion. The process of reconnection diffusion is based on the model of 3D reconnection 
of weakly turbulent magnetic fields and is applicable to both fully ionized and partially ionized gas.
The rate of reconnection diffusion does not depend on the level of ionization and therefore the
usually employed ambipolar diffusion idea gets irrelevant for magnetic field transport in turbulent fluids.
We claim that the reconnection diffusion process is a manifestation of the violation of flux conservation 
in highly conducting turbulent fluids. We discuss the consequences of reconnection diffusion for star formation
and stress. We show that reconnection diffusion on large scales is independent of small scale
magnetic field dynamics of magnetic fields. We conclude that numerical simulations correctly
represents the diffusion of actual astrophysical magnetic fields in flows with substantially larger
Lundquist numbers if these simulated regions regions are turbulent. 
\end{abstract}

\section{The problem of flux freezing}

Interstellar media are known to be turbulent and magnetized,  and both turbulence and magnetic field are important for star formation (see Armstrong et al. 1994, Chepurnov \& Lazarian 2009, Crutcher 2012). The existing star formation paradigm has been developed with the concept that flux freezing holds,
i.e. that the magnetic field is well coupled with ions and electrons in the media (Alfv\'{e}n 1942). Therefore it has been assumed assumed that the way to move matter across magnetic field lines should involve ambipolar diffusion (see Mestel 1965, 1966, Mouschovias et al. 2006). According to the theory, 
the change of the flux to mass ratio happens due to {\it ambipolar diffusion}, i.e. 
to the drift of neutrals which do not feel magnetic fields directly, but only through ion-neutral collisions. Naturally, in the presence of gravity, neutrals get concentrated towards the center of the gravitational potential while magnetic fields resist compression and therefore leave the
forming protostar. The rate of ambipolar diffusion for a cloud in gravitational equilibrium with the magnetic field 
depends only on the degree of ionization of the media. 

Magnetic fields are important at all stages of star formation. In many instances the ideas of star formation based exclusively on ambipolar diffusion have been challenged by observations (Troland \& Heiles 1986, Shu et al. 2006, Crutcher et al. 2009, 2010a, see Crutcher 2012 for a review).

While astrophysical fluids show a wide variety of properties in terms of their
collisionality, degree of ionization, temperature etc., they share a common
property, namely, most of the fluids are turbulent. The turbulent state of the
fluids arises from large Reynolds numbers $Re\equiv LV/\nu$, where $L$ is the
scale of the flow, $V$ is it velocity and $\nu$ is the viscosity, associated
with astrophysical media. Note, that the large magnitude of $Re$ is mostly the
consequence of the large astrophysical scales $L$ involved as well as the fact
that (the field-perpendicular) viscosity is constrained by the presence of
magnetic field.

The key concept of magnetic flux freezing has been challenged recently. On the basis of the model
of fast magnetic reconnection in Lazarian \& Vishniac (1999, henceforth LV99), Lazarian (2005) claimed that the removal of magnetic fields from turbulent plasma can happen during star formation due to magnetic reconnection rather than slow ambipolar drift (see also Lazarian \& Vishniac 2009). The process that was later termed {\it reconnection diffusion} does not depend on the degree of ionization, but rather on the properties of turbulence. The numerical confirmation of the idea was presented for molecular clouds and circumstellar accretion disks in Santos-Lima et al. (2010, 2012). The numerical testing of the LV99 reconnection model that underpins the concept of reconnection diffusion was successfully tested in Kowal (2009, 2012). Compared to this testing of reconnection diffusion, the testing in Kowal et al. (2009, 2012) was performed with much higher numerical resolution of the reconnection region and much better control of turbulence and other input parameters. In addition, more recent formal mathematical studies aimed at understanding of magnetic field dynamics in turbulent fluids supported the LV99 predictions (see Eyink, Lazarian \& Vishniac 2011). 

We note that the interstellar medium is collisional (Yamada et al. 2006), and therefore ideas of collisionless reconnection (see Shay \& Drake 1998, Shay et al. 1998, Bhattacharjee et al. 2005, 
Cassak et al. 2006) are not applicable.
 In addition, a theory of magnetic reconnection is necessary to understand whether
reconnection is represented correctly in numerical simulations. One should keep
in mind that reconnection is fast in computer simulations due to high numerical
diffusivity. Therefore, if there are situations where magnetic fields reconnect
slowly, numerical simulations do not adequately reproduce the astrophysical
reality. This means that if collisionless reconnection is indeed the only way to
make reconnection fast, then the numerical simulations of many astrophysical
processes, including those in interstellar media, which is collisional at the
relevant scales, are in error\footnote{Note, that it would be wrong to conclude
that reconnection must always be fast on the empirical grounds, as solar flares
require periods of flux accumulation time, which correspond to slow
reconnection.}.

To understand the difference between reconnection in astrophysical situations
and in numerical simulations, one should recall that the dimensionless
combination that controls the resistive reconnection rate is the Lundquist
number\footnote{The magnetic Reynolds number, which is the ratio of the magnetic
field decay time to the eddy turnover time, is defined using the injection
velocity $v_l$ as a characteristic speed instead of the Alfv\'en speed $V_A$,
which is taken in the Lundquist number.}, defined as $S = L_xV_A / \lambda$,
where $L_x$ is the length of the reconnection layer, $V_A$ is the Alfv\'en
velocity, and $\lambda=\eta c^2/4\pi$ is Ohmic diffusivity. Because of the large
astrophysical length-scales $L_x$ involved, the Lundquist numbers
are huge in astrophysics, e.g. for the ISM they are about $10^{16}$, while present-day MHD
simulations correspond to $S<10^4$. As the numerical efforts scale as $L_x^4$,
where $L_x$ is the size of the box, it is feasible neither at present nor in the
foreseeable future to have simulations with realistically Lundquist numbers.
As the brute approach fails, it is important to  understand if and when numerical
 correctly represent astrophysical reality in spite of the tremendous difference in 
 Lundquist numbers. 

\section{3D reconnection in weakly turbulent fluid}

If magnetic fields cannot reconnect fast, the intersection magnetic filed lines should create 3D elastic structure making
magnetized fluid Jello-like. This would automatically mean that numerical simulations of magnetized fluids
cannot represent astrophysical phenomena which Lundquist numbers are much larger than those
of the simulations.  To understand whether fluid-like behavior is possible we should consider the process of magnetic reconnection. 
\begin{figure}
\centering
 \includegraphics[height=.25\textheight]{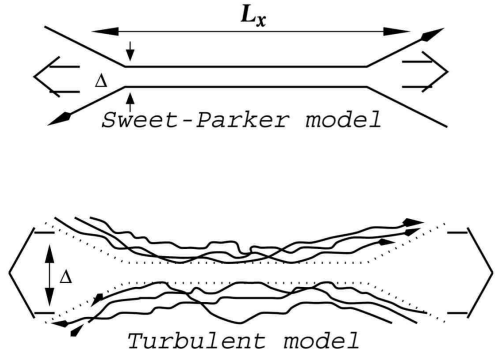}
 \caption{{\it Upper panel}. Sweet-Parker reconnection. $\Delta$ is limited by resistivity and is small. {\it Lower panel}: reconnection according to the LV99 model. $\Delta$ is determined by turbulent field wandering and can be large.  From Lazarian, Vishniac \& Cho (2004).}
\label{LV}
\end{figure}

The corresponding study of 3D magnetic reconnection in the presence of turbulence was performed in LV99. 
According to the latter study, magnetic reconnection of a turbulent field is {\it fast}, i.e. independent of resistivity and it happens at a rate
\begin{equation}
v_{rec, LV99}\approx V_A (L_x/L)^{1/2}M_A^2 ~~~{\rm for}~L_x<L,
\label{LV99}
\end{equation}
where $L_x$ is the horizontal extent of the reconnection zone (see Figure \ref{LV}), $L$ is the injection
scale of turbulence, $V_A$ is the Alfv\'{e}n velocity and $M_A\equiv V_L/V_A$ is the magnetic Mach number. In
Eq. (\ref{LV99}) it is assumed that turbulence is subAlfv\'{e}nic. For $L_x>L$, LV99 got the expression that differs from Eq.~(\ref{LV99}) by a change of the power $1/2$ to $-1/2$. 

The LV99 model of magnetic reconnection generalized the classical Sweet-Parker model of reconnection (see upper plot in Figure \ref{LV}) for the case
of turbulent fields.  The enhancement of the reconnection rate compared to the Sweet-Parker model
is achieved by increasing the outflow region $\Delta$ by accounting for magnetic field wandering. Indeed, in the Sweet-Parker model the reconnection speed is limited by the outflow of the matter from
the reconnection layer and $\Delta$ is determined by Ohmic diffusion. Therefore, 
\begin{equation}  
v_{rec}= V_A \frac{\Delta}{L_x} 
\label{vrec} 
\end{equation}
is much less than $V_A$  since $\Delta \ll L_x$. At the same time $\Delta$ becomes comparable with $L$
for $V\sim V_A$ provided that $\Delta$ is determined by the turbulent field wandering. 

\section{LV99 model and Richardson diffusion}

 At the scales less that the scale of injection of the strong MHD turbulence, i.e. for $l<l_{trans}$ (LV99, Lazarian 2006):
\begin{equation}
l_{trans}\sim L(V_L/V_A)^2\equiv LM_A^2.
\label{trans}
\end{equation}
 the magnetic field lines exhibit accelerated diffusion. Indeed, the separation between two particles $dl(t)/dt\sim v(l)$ for Kolmogorov turbulence is $\sim \alpha_t l^{1/3}$, where $\alpha$ is proportional to a cube-root of the energy cascading rate, i.e. $\alpha_t\approx V_L^3/L$ for turbulence injected with superAlvenic velocity
$V_L$ at the scale $L$. The solution of this equation is
\begin{equation} 
l(t)=[l_0^{2/3}+\alpha_t (t-t_0)]^{3/2},
\label{sol}
\end{equation}
which provides Richardson diffusion or $l^2\sim t^3$. The
accelerating character of the process is easy to understand physically. Indeed, the larger the separation 
between the particles, the faster the eddies that carry the particles apart. 
 
 Eyink, Lazarian \& Vishniac (2011) showed that the presence of Richardson diffusion induces fast magnetic reconnection and re-derived expressions
 for the reconnection rate from the Richarson diffusion concept. This provides a way to understand the applicability of LV99 approach to more
 involved cases, e.g. for the case of fluid with viscosity much larger than magnetic diffusivity, i.e. the high Prandtl number fluids. A partially ionized
 gas is an example of such a fluids, which is essential to understand in terms of star formation. 
 
In high Prandtl number media the GS95-type turbulent motions decay at the scale $l_{\bot, crit}$, which is much larger than the scale of at which Ohmic dissipation gets important. Thus over a range of scales less than $l_{\bot, crit}$ magnetic fields preserve their identity and are being affected by the shear on the scale $l_{\bot, crit}$. This is the regime of turbulence described in Cho, Lazarian \& Vishniac (2002) and Lazarian, Vishniac \& Cho (2004).  In view of the findings in Eyink et al. (2011) to establish when magnetic reconnection is fast and obeys the LV99 predictions one should establish the range of scales at which magnetic fields obey Richardson diffusion. It is easy to see that the transition to the Richardson diffusion happens when field lines get separated by the perpendicular scale of the critically damped eddies $l_{\bot, crit}$. The separation in the perpendicular direction starts with the scale $r_{init}$ follows the Lyapunov exponential growth with
the distance $l$ measured along the magnetic field lines, i.e.
$r_{init} \exp(l/l_{\|, crit})$, 
where $l_{\|, crit}$ corresponds to critically damped eddies with $l_{perp, crit}$. It seems natural to associate $r_{init}$ with the separation of the field lines arising from Ohmic resistivity on the scale of the critically damped eddies
\begin{equation}
r_{init}^2=\eta l_{\|, crit}/V_A,
\label{int}
\end{equation}
where $\eta$ is the Ohmic resistivity coefficient. 

In this formulation the problem of magnetic line separation is similar to the anisotropic analog of the Rechester \& Rosenbluth (1978) problem (see Narayan \& Medvedev 2003, Lazarian 2006) and therefore distance to be covered along magnetic field lines before the lines separate by the distance larger than the perpendicular scale of viscously damped eddies is equal to 
\begin{equation}
L_{RR}\approx l_{\|, crit} \ln (l_{\bot, crit}/r_{init})
\label{RR}
\end{equation}
Taking into account Eq. (\ref{int}) and that 
\begin{equation}
l_{\bot, crit}^2=\nu l_{\|, crit}/V_A,
\end{equation}
where $\nu$ is the viscosity coefficient. Thus Eq. (\ref{RR}) can be rewritten
\begin{equation}
L_{RR}\approx l_{\|, crit}\ln Pt
\label{RR2}
\end{equation}
where $Pt=\nu/\eta$ is the Prandtl number. . 

If the current sheets are much longer than $L_RR$, then magnetic field lines undergo Richardson diffusion and according to  Eyink et al. (2011) the reconnection follows the laws established in LV99. In other words, on scales significantly larger than the viscous damping scale LV99 reconnection is applicable. At the same time on scales less than $L_{RR}$ magnetic reconnection may be slow\footnote{Incidentally, this can explain the formation of density fluctuations on scales of thousands of Astronomical Units, that are observed in the ISM.}. Somewhat more complex arguments were employed in Lazarian et al. (2004) to prove that the reconnection is fast in the partially ionized gas. For our further discussion it is important that LV99 model is applicable both to fully ionized and partially ionized plasmas.

\section{Reconnection diffusion model}

The rate given by Eq. (\ref{LV99}) is sufficient to allow the magnetic field to disentangle during the turnover of the eddies. 
Such eddies are a part of the picture of strong Alfv\'{e}nic turbulence (see Goldreich \& Sridhar 1995, henceforth GS95)\footnote{The Alfv\'{e}nic cascade develops in compressible MHD turbulence
independently from the cascade of fast and slow modes, and its properties in compressible and incompressible
turbulence are very similar (Cho \& Lazarian 2003, Kowal \& Lazarian 2010). For our arguments related to reconnection diffusion the compressible motions are of secondary
importance. Indeed, the LV99 reconnection model is governed by field wandering induced by the 
Alfv\'{e}nic cascade and turbulent mixing and also depends on the
solenoidal component of the fluid.}. As a result, mixing eddy-type motions are possible in high Lundquist number fluids.

The peculiarity of reconnection diffusion is that it requires nearly parallel magnetic field lines to reconnect, while
the textbook description of reconnection frequently deals with anti-parallel description of magnetic field lines.
One should understand that the situation shown in Figure \ref{LV} is just a cross section of the magnetic fluxes
depicting the anti-parallel components of magnetic field. Generically, in 3D reconnection configurations the sheared 
component of magnetic field is present. This component is also frequently referred to as "guide field" in the
reconnection literature. 

\begin{figure}
\includegraphics[width=0.95\columnwidth,height=0.25\textheight]{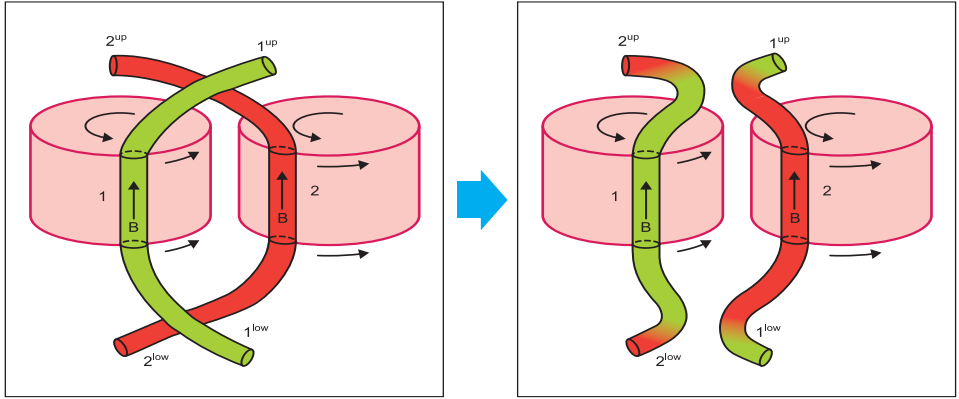}
\caption{Reconnection diffusion: exchange of flux with entrained matter. Illustration of the mixing of matter and
magnetic fields due to reconnection as two flux tubes of different eddies interact. Only one scale of turbulent
motions is shown. In real turbulent cascade such interactions proceed at every scale of turbulent motions.}
\label{mix}
\end{figure}

As we discuss below reconnection diffusion is closely connected with the reconnection between adjacent
Alfv\'{e}nic eddies (see Figure \ref{mix}). As a result, adjacent flux tubes exchange their segments with entrained plasmas
and flux tubes of different eddies get connected. This process involves eddies of all the sizes along the cascade and ensures
fast diffusion which has similarities with turbulent diffusion in ordinary hydrodynamic flows. 

\section{Theoretical expectations}

Ions and neutrals closely follows laminar magnetic field lines. Turbulent reconnection requires that magnetic flux
gets leaky and plasma move perpendicular to the mean magnetic field. In this situation the mass loading on 
magnetic field lines changes and the mass to flux ratio does not stay constant.

In the absence of gravity reconnection diffusion would tend to distribute magnetic flux and volume uniformly in 
the volume. Therefore any initial correlation of density and magnetic fields should dissipate due to reconnection
diffusion.

In the presence of gravity one should expect density to diffuse towards the gravitating center, while leaving magnetic
field behind. This would happen even in purely ionized plasmas, as reconnection diffusion does not depend on the 
degree of ionization. 

Plasma motions along magnetic field lines is expected to be diffusive due to magnetic field wandering. Therefore
the concepts of collection of matter along magnetic field lines must be modified in the presence of turbulence.

\section{Numerical testing of reconnection diffusion}

Numerical testing of reconnection diffusion concept have been performed in a series of papers by Santos-Lima et al. (2010,
2012). Both diffusion in the absence and in the presence of gravitational potential was tested. The results corresponded
well to the theoretical expectations. In particular, in Santos-Lima et al. (2012) the study of the formation of circumstellar
accretion disks has been performed. In the absence of turbulence the disks did not form, demonstrating the well-known 
magnetic breaking catastrophe behavior. Indeed, ambient interstellar magnetic field were getting stronger during the phase of accretion
and were very efficient in removing angular momentum. Reconnection diffusion, however, was removing magnetic fields from the
turbulent accreting matter and was shown to be able to resolve the magnetic breaking paradox. This corresponded to the theoretical
expectations in Lazarian \& Vishniac (2009). 

Numerical modeling of molecular clouds in Santos-Lima et al. (2010) demonstrated the ability of matter to concentrate towards the gravitational 
potential leaving magnetic field behind. The effect was shown to be present for different initial conditions.

 \section{Comparing reconnection diffusion predictions and observations}
 
 A comparison of the reconnection diffusion expectations with observations was performed in Lazarian, Esquivel \& Crutcher (2012). 
 In particular, results of Crutcher (2010) on the difference of the magnetization of cores and evelopes was analyzed. The traditional
 ambipolar diffusion requires that the less ionized and therefore more diffusive core should lose magnetic field faster than the 
 envelope that has larger degree of ioniziation. There a larger mass to flux ratio is expected in the core compared to the envelope.
 
 Contrary to the above expectations the observations indicated that the mass to flux ratio is larger in the envelope than in the core.
 This cannot be explained with ambipolar diffusion. At the same time, reconnection diffusion presents us with a number of possibilities.
 As the reconnection diffusion does not depend on the degree of ionization this removes the constraint that the magnetic field diffusion
 should happen faster out of the core. For instance, in a model of a homogeneous molecular cloud boadering with the diffuse media with lower 
 magnetization, the central part of the cloud keeps larger magnetization than its periphery. In the case of the core and the envelope system,
 the core with the lower level of turbulence and larger magnetic field may be subAlfenic, i.e. have the Alfven Mach number $M_A<1$ and 
 exhibit the reconnection diffusion that is suppressed by the factor $M_A^3<1$. Due to this factor, reconnection diffusion has an interesting
 property, namely, the larger initial magnetization, the slower the diffusion. Thus suggests that in the case of a collapse that compresses
 magnetic field the regions of higher magnetization will tend to lose magnetic flux slower, increasing the contrast between regions of low
 and high magnetization. This is provided that the regions have the same size. Otherwise, for the same degree of magnetization
 smaller regions will tend to lose the excess of magnetic field quicker. Indeed, the diffusion coefficient $\kappa$ scales as $l^{4/3}$, while 
 the rate of diffusion scales as $\kappa/l^2$ and therefore $\sim l^{-2/3}$. In the presence of self-gravity, however, one expect the denser
 regions to decouple from the cascade and the turbulent relation to be modified. making cores substantially less turbulent than envelopes. 
 
 All in all, while the decrease of the core size tends to make the diffusion out of the core faster, other effects, i.e. decrease of the turbulence
 in the core due to decoupling of the motions in the core from the outside cascade and smaller Alfven Mach number of the core tend to 
 prolong the time of the diffusion of magnetic flux out from the core. Due to the effects discussed above, the 
 inhomogeneity of the envelope may also contribute to the higher mass to flux  ratio observed in the envelope.
 
 An additional effect related to reconnection diffusion may be important for enhancing the mass to flux ratio in envelops compared to 
 cores. If the matter is accumulated along magnetic field lines due to external forcing rather than self-gravity, the matter will not follow
 the hour glass morphology of the field, but due to field wandering should preferentially reside in the envelope. 
 
 All these processes should be quantified better, but, it is clear that, unlike the process of ambipolar diffusion, reconnection diffusion 
 allow situations that the cores are more magnetized than envelops. Further research of the issue as well as more extensive
 comparison of the observational data with model predictions  are required.
 
 Lazarian, Esquivel \& Crutcher (2012) compared the rates of reconnection diffusion of magnetic fields out of the cloud with the rate of
 gravitational collapse and calculated the critical densities at which clouds should collapse trapping the magnetic flux. The results of these
 %calculations are illustrated by Figure \ref{diffvsff}. The column densities of the clouds at which the magnetic trapping happens were
  calculations are encouraging. The column densities of the clouds at which the magnetic trapping happens were
found to correspond to observations in Crutcher (2012). 
 
 %\begin{figure*}
%\centering
% \includegraphics[width=.40\textwidth]{figure9a.eps}
% \includegraphics[width=.40\textwidth]{figure9b.eps}
 %\caption{Illustration of the parameter space
% for which reconnection diffusion is faster than the gravitational collapse. The plots corresponds to three different models of turbulent injection described in the
% text. Lines corresponding to $l_{upper}$, $l_{lower}$ and the Jeans scale $R_J$ are shown. The
 %horizontal lines corresponds to the injection scale of $30$ pc.  {\bf Left panel}: Case of magnetic field of $B=6\mu\mathrm{G}$. {\bf Right panel}: 
 %Case of $B=30\mu\mathrm{G}$ magnetic field. }
%\label{diffvsff}
%\end{figure*} 
 
 It is encouraging that the reconnection diffusion concept provides new ways of approaching challenging problems of star formation in extreme environments.  For instance, galaxies emitting more than $10^{12}$ solar luminosities in the far-infrared are called ultra-luminous infrared galaxies or ULIRGs. The physical conditions in such galaxies are extreme with a very high density of 
cosmic rays (see Papadopoulos et al. 2011). Ambipolar diffusion is expected to be suppressed due to cosmic ray
ionization. At the same time these environments have the highest star formation rates. This is suggestive of a process that removes magnetic fields independent of the level of ionization. Reconnection diffusion does not depend on ionization and provides a natural explanation. 

It looks that reconnection diffusion can provide solutions to other long standing problems. For instance, the observed star formation rate is about the same in galaxies with low metallicities as in galaxies with high metallicities (see Elmegreen \& Scalo 2004). This is difficult to understand if magnetic field loss is governed by ambipolar diffusion, that is supposed to be much faster in low-$Z$ galaxies with high metallicities. As a result,  ambipolar diffusion theory predicts that the Initial Mass Function (IMF) and therefore the star formation rate would be shifted in low-$Z$ galaxies. This prediction contradicts to observations. As we discussed earlier, reconnection diffusion efficiency does not depend on metallicity, and this provides a solution consistent with observations. 
 
In addition, an important property of reconnection diffusion is that, unlike ambipolar diffusion, it not only provides the removal of the magnetic field, but also induces its turbulent mixing, which tends to make the distribution of magnetic field uniform. Thus turbulent mixing mediated by reconnection diffusion helps to keep the diffuse media in a magnetized subcritical state. In this situation, the external pressure is important for initiating collapse,  which well corresponds to the observations of numerous small dark clouds not forming stars in the inter-arm regions of galaxies (Elmegreen 2011).    

Finally, the reconnection diffusion concept implies that in realistic turbulent media there is no characteristic density for the collapse to be initiated. Therefore any cloud with the appropriate virial parameter (see McKee \& Zweibel 1992) can form stars. The difference between different clouds arises from the density controlling the timescale of the collapse and kinetic energy determined either by the level of steering by turbulence or thermal motions of particles. Directly, the requirement of clouds to be molecular is not present for the reconnection diffusion to induce star formation. The only difference between molecular and atomic clouds is that the former have lower temperatures.
 
 \section{Reconnection diffusion and MHD numerical simulations}
 
 Reconnection diffusion is
based on the numerically tested LV99 model of fast reconnection in turbulent media as well as more recent insights into the violation of the frozen-in condition in turbulent magnetized fluids (see Eyink et al. 2011). 
The direct 3D MHD simulations of the magnetic field diffusion in Santos-Lima et al. (2010, 2012) showed consistency with the theoretical expectations, but such
 simulations on their own cannot be used to justify the concept of reconnection diffusion. The interpretation of the results requires the proper understanding of scaling of magnetic reconnection with the dimensionless combination called the Lunquist number $S\equiv (L_{cur. sh} V_A/\eta)$, where $L_{cur. sh.}$ is the
extent of the relevant current sheet, $\eta$ is Ohmic diffusivity. The Lundquist numbers in molecular clouds and in the corresponding simulations
differ by a factor larger than $10^5$. In this situation one can establish the correspondence between the numerical simulations and astrophysical reality
only if the reconnection does not depend on $S$. The independence from $S$ of magnetic reconnection is the conclusion of LV99 model. This model
has been tested in Kowal et al. (2009, 2012) via a set of dedicated numerical simulations that confirmed the scaling predictions in LV99. This work
exemplifies the advantages of the synergy of scaling arguments with numerical simulations as opposed to the ``brute force numerical approach'', which may not be productive while dealing with turbulence.

One can argue that the process of reconnection diffusion was present in some of high resolution numerical simulations, although the researchers did not identify
 the process. However, without clear identification of the role of turbulence in fast reconnection, one may not be sure when the results are due to physically motivated reconnection diffusion and when they are
the consequence of the bogus effects of numerical diffusion. For instance, Crutcher, Hakobian \& Troland (2009) refer to the simulations in Luntilla et al. (2009) that produce, in 
agreement with observations, higher magnetization of the cloud cores. If these cores are of the size of several grid units across, numerical
effects rather than reconnection diffusion may be dominant and turbulence is suppressed at these scales.

There are some very good news for numerical simulations, however. The LV99 model predicts that the reconnection rates in turbulent fluids are independent of the local physics, but are determined by the turbulent motions. Therefore parasitic numerical effects that induce poorly controlled small scale diffusivity of magnetic field lines are not important on the scales of turbulent
motions. In other words, the low resolution numerics may provide an adequate representation of high Lundquist
number astrophysical turbulence as far as the reconnection diffusion is concerned. 
 
 On the contrary, if the structures studied in numerical simulations
(e.g. cores, filaments, shells) lose turbulence due to numerical diffusivity effects, we predict that the diffusion of magnetic flux in those simulated structures differs significantly from the diffusion in the actual interstellar structures where turbulence persists. As a result, naive convergence studies based on increasing the numerical resolution several times cannot notice the problem unless the resolution increased to the degree that the aforementioned
structures become turbulent.

\section{Reconnection diffusion and simulations with ambipolar drift}

A number of studies attempted to understand the role of joint action of turbulence and ambipolar diffusion. For instance, Heitsch et al. (2004, henceforth HX04) performed 2.5D simulations of turbulence with two-fluid code and examined the decorrelation of neutrals and magnetic field in the presence of turbulence.
 The study reported an enhancement of diffusion rate compared
to the ambipolar diffusion in a laminar fluid. HX04 correctly associated the enhancement with turbulence
creating density gradients that are being dissolved by ambipolar diffusion (see also Zweibel 2002). However,
in 2.5D simulations of HX04 the numerical set-up artificially precluded reconnection from taking place as magnetic
field was perpendicular to the plane of 2D mixing and therefore magnetic field lines were absolutely parallel to each other. This will not 
happen in any realistic astrophysical situation where reconnection will be an essential part of the physical picture. Therefore
we claim that a treatment of "turbulent ambipolar diffusion" without addressing the reconnection issue is of academic interest.

Incidentally, the
authors of HX04 reported an enhanced rate that is equal to the turbulent diffusion rate $L V_L$, which is the
result expected within the reconnection diffusion picture for the special set-up studied. The fact that 
ambipolar diffusion rate does not enter the result in HX04 suggests that ambipolar diffusion is irrelevant for the diffusion of matter in the presence 
of turbulence. This is another reason not to call the observed process "turbulent ambipolar diffusion"\footnote{A similar process takes place in the case of molecular diffusivity in turbulent hydrodynamic flows. The result for the latter flows is well known: in the turbulent regime, molecular diffusivity is irrelevant
for the turbulent transport. The process is not called therefore "turbulent diffusivity" without adding word "molecular".}.

Therefore we believe that HX04 captured in their simulations a special degenerate case of 2.5D turbulent diffusion where due to a special set up the
reconnection is avoided and magnetic field lines do not intersect. We also note that, in the presence of turbulence, the independence of the gravitational collapse from the ambipolar diffusion rate was reported in numerical simulations by Balsara, Crutcher \& Pouquet (2001). 

\section{Summary}

~~~~~~~~~1. Magnetic reconnection and turbulence are inter-dependent processes. Magnetic reconnection gets fast in the presence of turbulence, while MHD turbulence
requires magnetic reconnection to develop eddy-type structures.

2. Turbulence in the presence of magnetic reconnection induces diffusion process that we term "reconnection diffusion" to stress the importance of magnetic
reconnection for transporting magnetic fields and matter. 

3. Reconnection diffusion in most cases is the process that is faster than ambipolar diffusion. Therefore it is expected to dominate the removal of magnetic fields
from the forming molecular clouds.

4. Reconnection diffusion acts both in fully ionized and partially ionized gas. The effects of ambipolar diffusion are important only at the small scales and do not affect the important large scale dynamics of magnetic field diffusion.

5. As diffusion of magnetic fields and matter do not depend on the microphysical processes turbulent volumes of numerical simulations represent the magnetic
diffusion correctly in spite of tremendous differences in the Lundquist number of simulations and astrophysical reality.

%%==============================================================================
%%
\paragraph{Acknowledgements.}

The research is supported
by the Center for Magnetic Self-Organization in Laboratory and Astrophysical
Plasmas, the Vilas Associate Award and the NSF Grant AST-1212096. Hospitality of
the International Institute of Physics (Natal, Brazil) is acknowledged.

\end{document}